\documentclass[prc,twocolumn,twoside,preprintnumbers,superscriptaddress,nofootinbib,letterpaper]{revtex4-2}

\usepackage{amsmath,slashed}
\usepackage{graphicx, graphics, color}
\usepackage{xspace}
\usepackage{enumerate}
\usepackage{varwidth}
\usepackage{microtype}

\usepackage{physics}
\usepackage{amssymb}
\usepackage{mathtools}
\usepackage{dsfont}
\usepackage{multirow}
\usepackage{booktabs}

\usepackage[colorlinks=true,citecolor=blue,linkcolor=blue,urlcolor=blue]{hyperref}

\usepackage{orcidlink}

\newcommand{\prlsection}[1]{\textit{#1.---}}
\newcommand{\iso}[2]{\ensuremath{^{#2}\mathrm{#1}}}
\newcommand{\MeV}{\ensuremath{\mathrm{MeV}}}
\newcommand{\keV}{\ensuremath{\mathrm{keV}}}
\newcommand{\fm}{\ensuremath{\mathrm{fm}}}
\newcommand{\epssplit}{\ensuremath{\varepsilon_\mathrm{split}}}

\begin{document}

\title{Computational schemes for the Magnus expansion\\ of the in-medium similarity renormalization group}
\author{Matthias Heinz\,\orcidlink{0000-0002-6363-0056}}
\email{heinzmc@ornl.gov}
\affiliation{National Center for Computational Sciences, Oak Ridge National Laboratory, Oak Ridge, TN 37831, USA}
\affiliation{Physics Division, Oak Ridge National Laboratory, Oak Ridge, TN 37831, USA}

\begin{abstract}

The in-medium similarity renormalization group (IMSRG) is a popular many-body method used for computations of nuclei.
It solves the many-body Schrödinger equation through a continuous unitary transformation of the many-body Hamiltonian.
The IMSRG transformation is typically truncated at the normal-ordered two-body level,
the IMSRG(2),
but recently several approaches have been developed to capture the effects of normal-ordered three-body operators, the IMSRG(3).
In particular, a factorized approximation to the IMSRG(3) proposes to capture the leading effects of three-body operators at the same computational cost as the IMSRG(2) approximation.
This approach often employs an approximate scheme for solving the IMSRG equations,
the so-called hunter-gatherer scheme.
In this work, I study the uncertainty associated with this scheme.
I find that the hunter-gatherer scheme differs by up to $7\,\mathrm{MeV}$ for ground-state energies and $0.5\,\mathrm{MeV}$ for excitation energies from standard IMSRG(2) approaches.
These differences are in some cases comparable to the expected size of IMSRG(3) corrections.

\end{abstract}

\maketitle

\prlsection{Introduction}
Ab initio computations of nuclei
start from nuclear forces from chiral effective field theory
and solve the many-body Schrödinger equation using exact or systematically improvable methods~\cite{Hergert2020FP_AbInitioReview}.
Such calculations
have advanced tremendously,
with reproduction of a broad range of nuclear properties (including ground-state energies, spectra, nuclear radii, magnetic moments, electroweak transition strengths, and nuclear responses) across the nuclear chart~\cite{Morris2018PRL_Sn100, 
Gysbers2019NP_BetaDecay2BC, Soma2020PRC_SCGFlnl, King2020wmp, Stroberg2021PRL_AbInitioLimits, Hu2022NP_Pb208, Tichai2024PLB_BCC, Miyagi2024zvv, Porro2024tzt, Miyagi2025PLB_Rch4, Bonaiti2025arxiv_Pb266, Bonaiti2025txe}.
Ab initio calculations are now focusing on experimentally inaccessible properties,
including nuclear structure effects relevant for searches for new physics beyond the standard model~\cite{Gazda2017PRD_DMNuclResponses, Andreoli2019etf, Payne2019wvy, Novario2020dmr, Hu2022PRL_DMResponses, Glick-Magid2022uwb, Cirigliano2024msg, Belley2024lec, Farren2024spl,  Gennari2025sbn, Door2025PRL_YbBoson, Heinz2025PLB_MuToE, Ng2025hgx, King2025fph}
and masses and half lives of $r$-process nuclei~\cite{Li2025exk, Kuske2025tsm}.
For such predictions, it is essential to understand and quantify all uncertainties, e.g., due to truncations in the effective field theory or in the many-body method.

The in-medium similarity renormalization group (IMSRG)~\cite{Tsukiyama2011PRL_IMSRG, Hergert2016PR_IMSRG, Stroberg2017PRL_VSIMSRG, Stroberg2019ARNPS_VSIMSRG} is a systematically improvable many-body method
that performs a unitary transformation of the Hamiltonian
\begin{equation}
    \label{eq:UnitaryTransformation}
    H(s) = U(s)\,H\, U^\dagger(s)
\end{equation}
by solving the flow equation
\begin{equation}
    \label{eq:FlowEquation}
    \frac{dH(s)}{ds} = [\eta(s), H(s)]
\end{equation}
from $s=0$ towards $\infty$.
The generator $\eta(s)$ is chosen to generate the desired unitary transformation:
One either decouples the ground state from its excitations in the single-reference IMSRG~\cite{Tsukiyama2011PRL_IMSRG},
or one decouples a core and valence space from the rest of the Hilbert space in the valence-space IMSRG (VS-IMSRG)~\cite{Stroberg2017PRL_VSIMSRG}.

The IMSRG is typically solved in the IMSRG(2) approximation,
where all operators are truncated at the normal-ordered two-body level.
This approximation may be systematically improved by including normal-ordered three-body operators,
the IMSRG(3)~\cite{Heinz2021PRC_IMSRG3, Stroberg2024PRC_IMSRG3, Heinz2025PRC_Calcium}.
IMSRG(3) calculations are several orders of magnitude more expensive than IMSRG(2) calculations,
making them far from routine.
Still, the uncertainty due to the IMSRG(2) must be quantified for a complete uncertainty estimate.
To this end, Ref.~\cite{He2024PRC_IMSRG3f2} introduced a factorized approximation to the IMSRG(3),
the IMSRG(3f$_2$).
The IMSRG(3f$_2$) captures leading IMSRG(3) corrections at the same computational cost as the IMSRG(2),
providing an accessible way to obtain more precise IMSRG predictions
and quantify the many-body method uncertainty.

The IMSRG(3f$_2$) typically uses an approximate computational scheme for the IMSRG(2),
the hunter-gatherer scheme.
In this article, I study and quantify the uncertainty associated with the hunter-gatherer scheme.

\prlsection{Magnus expansion schemes}
The IMSRG may be solved by directly integrating Eq.~\eqref{eq:FlowEquation} from $s=0$ towards $\infty$.
Consistently transforming other operators to evaluate their expectation values or transition matrix elements requires solving additional flow equations in parallel:
\begin{equation}
    \label{eq:OpFlowEquation}
    \frac{dO(s)}{ds} = [\eta(s), O(s)]\,,
\end{equation}
for a given operator $O$.
This makes direct integration of Eqs.~\eqref{eq:FlowEquation} and~\eqref{eq:OpFlowEquation} prohibitively expensive when considering more than just a few operators.
The Magnus expansion~\cite{Morris2015PRC_Magnus}
directly parametrizes the unitary transformation
\begin{equation}
    U(s) = e^{\Omega(s)}
\end{equation}
with the anti-Hermitian Magnus operator $\Omega$.
One instead solves a flow equation for the Magnus operator:
\begin{equation}
    \label{eq:MagnusExpansion}
    \frac{d\Omega(s)}{ds} = \eta(s) + \frac{1}{2}[\Omega(s), \eta(s)] + \frac{1}{12}\big[\Omega(s), [\Omega(s), \eta(s)] \big] + \dots.
\end{equation}
The transformed Hamiltonian is then computed using the Baker-Campbell-Hausdorff (BCH) formula
\begin{equation}
    \label{eq:BCHExpansion}
    H(s) = H + [\Omega(s), H] + \frac{1}{2}\big[\Omega(s), [\Omega(s), H]\big] + \dots ,
\end{equation}
and any other operator may be transformed in the same way.
The Magnus expansion has the clear benefit that the unitary transformation is solved for only once
and then easily applied to any other operator.

An important subtlety is the role of the many-body approximation.
The Magnus approach and the direct integration of the flow equation~\eqref{eq:FlowEquation} (hereafter referred to as the flow equation approach)
are exactly equivalent only when the Magnus expansion in Eq.~\eqref{eq:MagnusExpansion} converges
and the commutators in Eqs.~\eqref{eq:FlowEquation}, \eqref{eq:MagnusExpansion}, and~\eqref{eq:BCHExpansion} are evaluated exactly.
In practice, all commutators are truncated.
In the IMSRG(2), this truncation is at the normal-ordered two-body level.
As a result, the Magnus approach and the flow equation approach
will differ at the level of the normal-ordered two-body approximation.

To mitigate this,
a modified computational scheme for the Magnus approach has been standardized~\cite{Stroberg2024PRC_IMSRG3},
referred to here as the split Magnus approach.
The unitary transformation is split into small successive  transformations:
\begin{align}
    U(s) &=U_{N+1}(s -s_N) \cdots U_2(s_2 -s_1)U_1(s_1) \nonumber \\
    &=e^{\Omega_{N+1}(s - s_N)}\cdots e^{\Omega_2(s_2 - s_1)}e^{\Omega_1(s_1)}\,.
\end{align}
The first transformation on the right transforms from $s=0$ to $s_1$,
the next transforms from $s=s_1$ to $s_2$,
and so on.
The splitting is decided based on the Frobenius norm 
of $\Omega_i(s - s_{i-1})$;
one chooses a splitting threshold $\epssplit$
and once the calculation reaches $s=s_i$ where
$||\Omega_i(s - s_{i-1})|| \geq \epssplit$
then $\Omega_i(s_i - s_{i-1})$ is saved
and a new unitary transformation is started from $s=s_i$,
$U_{i+1}(s - s_i) = e^{\Omega_{i+1}(s - s_i)}$.

This approach has a few benefits.
First, the norm of $\Omega_i$ remains small,
so the nested commutator expansions in Eqs.~\eqref{eq:MagnusExpansion} and~\eqref{eq:BCHExpansion}
converge quickly.
This reduces number of commutators evaluated to solve the IMSRG,
reducing the computational cost.
Second,
one can see that as $\epssplit \to 0$
it approaches the flow equation solution.
Each $\Omega_i$ performs an infinitesimal transformation of the Hamiltonian.
The integration from $s_i$ to $s_i + \Delta s$ in Eq.~\eqref{eq:MagnusExpansion} becomes $\frac{d\Omega_{i+1}(s - s_i)}{ds} = \eta(s_i)$,
yielding $\Omega_{i+1}(\Delta s) = \Delta s\, \eta(s_i)$.
Plugging this into Eq.~\eqref{eq:BCHExpansion} gives
\begin{equation}
    H(s_i + \Delta s) = H(s_i) + \Delta s [\eta(s_i), H(s_i)] + \mathcal{O}[(\Delta s)^2]\, .
\end{equation}

It is clear that this is a systematic way to approximate the flow equation approach using the Magnus expansion
at the price of storing several Magnus operators.
In practice, one chooses a value for $\epssplit$
so that the IMSRG transformation is close to the $\epssplit \to 0$ limit
but the number of Magnus operators stored remains moderate (e.g., $< 50$).
To store fewer Magnus operators,
one can increase the value for $\epssplit$.

\begin{figure}
    \centering
    \includegraphics{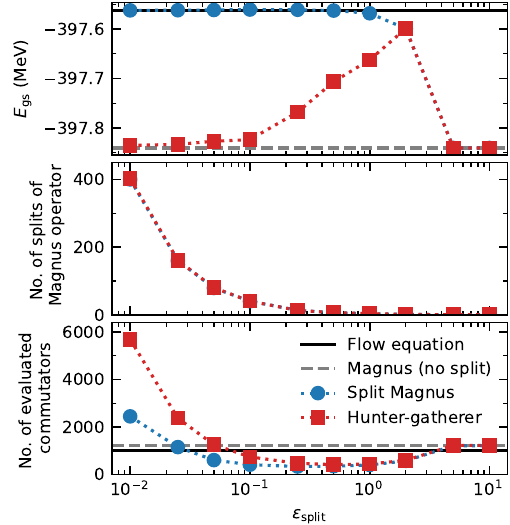}
    \caption{
    	Comparison of split Magnus and hunter-gatherer approaches for varying splitting thresholds \epssplit{}.
	The flow equation, Magnus (without splitting),  split Magnus, and hunter-gatherer results are indicated by the black, gray, blue, and red lines, respectively.
	I consider IMSRG(2) computations of \iso{Ca}{48} in a model space of seven major oscillator shells with frequency $\hbar\omega=16\,\MeV$.
    The top, middle, and bottom panels show the computed ground-state energies, number of Magnus operator splits, and number of evaluated commutators, respectively, for each of the calculations.
    }
    \label{fig:MethodComparison}
\end{figure}

The hunter-gatherer scheme avoids storing many Magnus operators by handling the splitting differently~\cite{Stroberg2024PRC_IMSRG3}.
The unitary transformation is split in two parts,
\begin{equation}
    U(s) = e^{\Omega_H(s)} e^{\Omega_G(s)}\,,
\end{equation}
the hunter parametrized by $\Omega_H(s)$,
which remains small,
and the gatherer parametrized by $\Omega_G(s)$.
When $||\Omega_H(s)||\geq \epssplit$,
the hunter is absorbed into the gatherer using the BCH expansion
and is set to 0.
One then continues to solve the flow equation for the Magnus operator until the the splitting threshold is reached.
The unitary transformation remains compact, requiring only two Magnus operators.

\begin{figure*}
    \centering
    \includegraphics{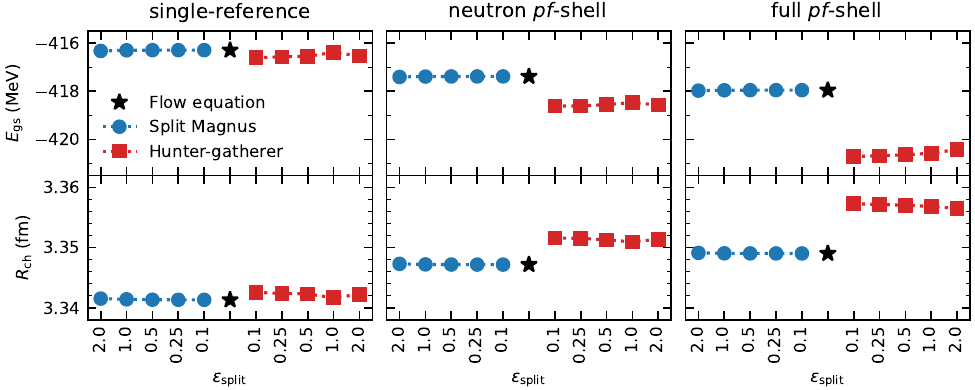}
    \caption{
    	Comparison of ground-state energies and charge radii of $^{48}$Ca
	predicted using the flow equation (black stars), split Magnus (blue circles),
	and hunter-gatherer (red squares) approaches
	for varying \epssplit{}.
	I consider single-reference IMSRG(2) calculations (left)
	and VS-IMSRG(2) calculations using 
	neutron $pf$-shell and full $pf$-shell valence spaces (middle and right, respectively).
    }
    \label{fig:Ca48Comparison}
\end{figure*}

Let us consider the limiting case of $\epssplit \to 0$
for the hunter-gatherer approach.
Following a split at $s_i$, $\Omega_H(s_i) = 0$,
and only $\Omega_G(s_i)$ parametrizes a nontrivial unitary transformation.
Integrating from $s_i$ to $s_i + \Delta s$, $\Omega_H(s_i + \Delta s) = \Delta s \eta(s_i)$.
Because \epssplit{} is small, the hunter is immediately absorbed into the gatherer:
\begin{equation}
    e^{\Omega_G(s_i + \Delta s)} = e^{\Delta s \eta(s_i)}e^{\Omega_G(s_i)}\,.
\end{equation}
Applying the BCH formula yields
\begin{align}
    \Omega_G(s_i + \Delta s) &= \Omega_G(s_i) + \Delta s \eta(s_i) + \Delta s \frac{1}{2} [\Omega_G(s_i), \eta(s_i)] \nonumber
    \\ & \quad
    + \Delta s \frac{1}{12} \big[\Omega_G(s_i), [\Omega_G(s_i), \eta(s_i)]\big] + \dots \nonumber 
    \\ & \quad 
    + \mathcal{O}[(\Delta s)^2],
\end{align}
recovering the flow equation for a single Magnus operator~\eqref{eq:MagnusExpansion}.
Clearly the hunter-gatherer scheme approaches the standard Magnus approach (without splitting) for small \epssplit{}.

In Fig.~\ref{fig:MethodComparison},
I demonstrate the differences between the split Magnus and hunter-gatherer approaches,
comparing against the two limiting cases, the flow equation approach and the normal Magnus approach without splitting.
In this case, the ground-state energy computed using the flow equation approach (black line)
differs from that computed using the normal Magnus approach (dashed gray line)
by $280\,\keV$.
For large values of \epssplit{},
the Magnus operator is never actually split,
so both the split Magnus and hunter-gatherer schemes are equivalent to the normal Magnus approach.
As \epssplit{} is decreased,
the Magnus operator is split increasingly often.
The split Magnus approach quickly reaches excellent agreement with the flow equation approach for $\epssplit = 0.5$--$1.0$.
The hunter-gatherer approach instead converges towards the normal Magnus result.
This convergence is much slower, requiring $\epssplit < 0.1$
and splitting the Magnus operator more than 75 times
to reach good agreement with the normal Magnus result.
For such small values of \epssplit{},
the total number of commutators evaluated
(which determines the computational cost of the calculation)
grows significantly in both the split Magnus and hunter-gatherer schemes.
The hunter-gatherer scheme requires more commutator evaluations
because the gatherer is large and the BCH expansions converge more slowly as a result.

The split Magnus approach offers a generous window of \epssplit{} values
where observables and computational costs are stable with respect to varying this parameter.
On the other hand,
for the hunter-gatherer scheme
in the window where the ground-state energy is stable with respect to \epssplit{} variation,
the computational cost varies by nearly an order of magnitude
(744 commutators for $\epssplit = 0.1$ versus 5687 for $\epssplit = 0.01$).
Clearly identifying a splitting threshold that is small enough to give accurate results, but not so small that the computational cost is prohibitive
is challenging in the hunter-gatherer approach.

\prlsection{The hunter-gatherer uncertainty}
I compare the hunter-gatherer and split Magnus schemes in IMSRG(2) and VS-IMSRG(2) calculations of the ground-state energy and charge radius of \iso{Ca}{48} in Fig.~\ref{fig:Ca48Comparison}.
Throughout, I use the 1.8/2.0~(EM) Hamiltonian~\cite{Hebeler2011PRC_SRG3NFits},
employ a model space of 13 major harmonic oscillator shells with $\hbar\omega=16\,\MeV$
and start from a Hartree-Fock reference state.
Three-body forces are truncated with $E_\mathrm{3max}=24$~\cite{Miyagi2023EPJA_NuHamil} throughout.
Only the method of solving the IMSRG(2) equations
and the splitting parameter \epssplit{} (where relevant) are varied.

On the left, I consider single-reference IMSRG calculations.
The results obtained by solving the IMSRG(2) via direct integration of the flow equation are indicated by black stars.
In blue, I show results obtained using the split Magnus approach for $\epssplit = 2.0$, 1.0, 0.5, 0.25, 0.1.
Overall, in all cases, the flow equation result is well approximated,
with, e.g., the ground-state energy differing only by $22\,\keV$ for $\epssplit = 2.0$.
For smaller \epssplit{}, the split Magnus results systematically converge towards the flow equation results.
This is not the case for the hunter-gatherer approach, which approximates the Magnus approach without splitting.
For all \epssplit{}
the predicted ground-state energies and charge radii differ visibly from the flow equation predictions.
In this case,
the deviations of $\approx 300\,\keV$
and $\approx 0.0015\,\fm$ are still considerably smaller
than the expected IMSRG(3) corrections
(estimated to be $\approx 3\,\MeV$ and $\approx 0.03\,\fm$ in Ref.~\cite{Heinz2025PRC_Calcium}).

\begin{figure}
    \centering
    \includegraphics{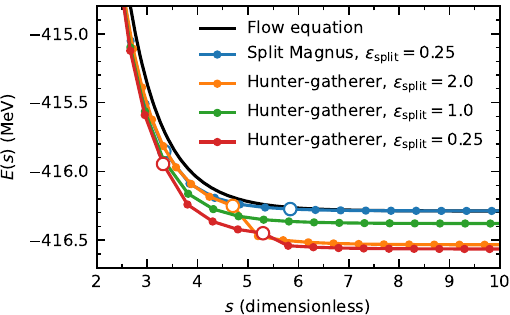}
    \caption{
    	Energies from the IMSRG(2) as a function of the flow parameter $s$.
	I compare direct integration of the flow equation (black line)
	with the split Magnus scheme (blue)
	and the hunter-gatherer scheme with splitting threshold $\epssplit = 2.0$, $1.0$, $0.25$ (orange, green, and red, respectively).
	Small circles indicate the integration steps,
	and large open circles indicate integration steps where the Magnus operator is split.
    }
    \label{fig:flow}
\end{figure}

One also sees that the predictions in the hunter-gatherer approach exhibit nontrivial dependence on \epssplit{}.
In Fig.~\ref{fig:flow} I investigate this further by considering the ground-state energy $E(s)$ over the integration in $s$.
Recall that $E(s\to\infty)$ gives the final ground-state energy.
In black, I show $E(s)$ obtained from the flow equation approach.
The split Magnus approach in blue (with $\epssplit = 0.25$)
closely follows the flow equation and reaches the same value
at $s\to\infty$.
The smaller dots indicate the integration steps,
and the large open circles are steps where the Magnus operator is split.
In this approach the splitting of the Magnus operator does not lead to discontinuities in $E(s)$
or cause any other problems during the integration.
In the hunter-gatherer approach,
$E(s)$ jumps when the Magnus operator is split
and the hunter is absorbed into the gatherer.
The new gatherer fails to generate the same transformation as the previous hunter and gatherer combined,
producing a discontinuity in the energy after the split.
This jump can be large:
For instance for $\epssplit = 2.0$,
the final split at $s=4.7$ causes $E(s)$ to suddenly decrease by $200\,\keV$.
The accumulated error from these jumps produces the difference between the flow equation and hunter-gatherer approaches.

One also sees that for larger values of \epssplit{} the hunter-gatherer approach is sensitive to where the splitting of the Magnus operator actually occurs.
For $\epssplit=1.0$,
the last split of the Magnus operator occurs around $s=1.9$ (not shown), so the final integration to $s\to\infty$ is stable.
For $\epssplit = 2.0$,
before the last split at $s=4.7$ the energy is still quite close to the standard split Magnus value;
after the split it jumps to be significantly lower.
This is in contrast to the split Magnus approach,
which is only weakly sensitive to the value of \epssplit{} and systematically converges to the flow equation result.
The hunter-gatherer scheme becomes less sensitive to the splitting for small values of \epssplit{},
consistent with the findings of Ref.~\cite{He2024PRC_IMSRG3f2}.

Returning to Fig.~\ref{fig:Ca48Comparison}, in the middle panels I consider instead VS-IMSRG(2) calculations using a neutron $pf$-shell valence space.
The predicted ground-state energy and charge radius from the flow equation approach differ from the single-reference IMSRG(2) values
because of the different decoupling used in the VS-IMSRG.
The two stage decoupling (first decoupling the core, and second decoupling the valence space) is more involved than the single-reference decoupling.
The resulting difference between the single-reference and neutron $pf$-shell results is due to the IMSRG(2) truncation.
One sees that the split Magnus approach again accurately reproduces the flow equation result.
The hunter-gatherer approach differs significantly,
by over $1.2\,\MeV$ for the ground-state energy
and over $0.004\,\fm$ for the charge radius.
The difference on the ground-state energy is of the same size as the expected VS-IMSRG(3) correction,
estimated to be $\approx 2\,\MeV$~\cite{Heinz2025PRC_Calcium}.

In the right panels, I consider VS-IMSRG(2) calculations using a full $pf$-shell valence space.
This valence space is large, and the VS-IMSRG(2) decoupling is more difficult.
The split Magnus approach accurately approximates the flow equation result,
while the hunter-gatherer approach differs by over $2\,\MeV$ for the ground-state energy and over $0.008\,\fm$ for the charge radius.

\begin{table}
    \caption{
        \label{tab:ComparisonTable}
        Comparison of VS-IMSRG(2) predictions using the split Magnus and hunter-gatherer schemes
        (using $\epssplit = 0.25$)
        for selected observables in  \iso{C}{12}, \iso{Al}{27}, \iso{Ca}{48}, and \iso{Bi}{209}.
        Energies are given in MeV, radii are given in fm, and magnetic moments are given in units of the proton magneton $\mu_N$.
        All calculations use a model space of 13 major oscillator shells with a frequency of $\hbar\omega=16\,\MeV$,
        except \iso{Bi}{209} where I use $\hbar\omega=12\,\MeV$.
        The model-space truncations are not sufficient for converged predictions in \iso{Bi}{209},
        but a meaningful comparison between the approaches can still be made.
        I employ the following valence spaces:
        $p$ shell for \iso{C}{12};
        $sd$ shell for \iso{Al}{27};
        $pf$ shell for \iso{Ca}{48};
        and a valence space consisting of a \iso{Pb}{208} core with proton $1f_{7/2}$, $2p_{3/2}$, $2p_{1/2}$, $1f_{5/2}$, $0h_{9/2}$, $0i_{13/2}$ valence orbitals
        for \iso{Bi}{209}.
        The magnetic dipole moment of \iso{Bi}{209} is evaluated using only one-body currents.
    }
    \begin{ruledtabular}
    \begin{tabular}{lcrr}
    System & Observable & Split Magnus & Hunter-gatherer \\
    \\[-2.5ex] \hline & \\[-2.5ex]
    \iso{C}{12} & $E_\mathrm{gs}$ & $-93.8$ & $-94.3$ \\
    & $R_\mathrm{ch}$ & $2.473$ & $2.476$ \\
    & $E_\mathrm{ex}(2^+_1)$ & $5.1$ & $5.2$ \\
    \\[-2.5ex] \hline & \\[-2.5ex]
    \iso{Al}{27} & $E_\mathrm{gs}$ & $-225.2$ & $-226.9$ \\
    & $R_\mathrm{ch}$ & $2.994$ & $2.970$ \\
    & $E_\mathrm{ex}(1/2^+_1)$ & $0.4$ & $0.2$ \\
    \\[-2.5ex] \hline & \\[-2.5ex]
    \iso{Ca}{48} & $E_\mathrm{gs}$ & $-418.0$ & $-420.7$ \\
    & $R_\mathrm{ch}$ & $3.349$ & $3.357$ \\
    & $E_\mathrm{ex}(2^+_1)$ & $4.8$ & $4.6$ \\
    \\[-2.5ex] \hline & \\[-2.5ex]
    \iso{Bi}{209} & $E_\mathrm{gs}$ & $-1635.8$ & $-1642.9$ \\
    & $R_\mathrm{ch}$ & $5.223$ & $5.230$ \\
    & $\mu$ & 2.79 &  2.83 \\
    & $E_\mathrm{ex}(7/2^-_1)$ & $2.3$ &  $2.0$ \\
    & $E_\mathrm{ex}(13/2^+_1)$ & $3.8$ &  $3.3$ \\
    \end{tabular}
    \end{ruledtabular}
\end{table}

\begin{table}
    \caption{
        \label{tab:Bi209Table}
        Energies of states (in MeV) of \iso{Bi}{209} relative to the \iso{Pb}{208} ground state
        computed in the split Magnus and hunter-gatherer schemes (see Table~\ref{tab:ComparisonTable} for details).
        Results from coupled-cluster theory [PA-EOM-CCSD(3p-2h)$_\mathrm{pert}$]
        from Ref.~\cite{Bonaiti2025arxiv_Pb266}
        are provided for comparison.
    }
    \begin{ruledtabular}
    \begin{tabular}{lrrr}
    State & Coupled cluster & Split Magnus & Hunter-gatherer \\
    \\[-2.5ex] \hline & \\[-2.5ex]
    $9/2^-_1$ & $-5.14$ & $-5.21$ & $-4.97$ \\
    $7/2^-_1$ & $-2.77$ & $-2.87$ & $-2.98$ \\
    $13/2^+_1$ & $-1.12$ & $-1.37$ & $-1.63$ \\
    \end{tabular}
    \end{ruledtabular}
\end{table}

The difference between the flow equation and the hunter-gatherer approach systematically grows with the size of the valence space.
Single-reference IMSRG(2) calculations are reliably approximated by both the split Magnus and hunter-gatherer approaches.
For larger valence spaces, the hunter-gatherer approach is a significant approximation
that can change predicted observables by amounts that are comparable to the size of IMSRG(3) corrections (e.g., for energies and spectra~\cite{Heinz2025PRC_Calcium}).
This is summarized for selected observables in \iso{C}{12}, \iso{Al}{27}, \iso{Ca}{48}, and \iso{Bi}{209} in Table~\ref{tab:ComparisonTable}.
The hunter-gatherer scheme induces only small differences for \iso{C}{12},
with a small $p$-shell valence space,
while the differences induced for \iso{Al}{27}, \iso{Ca}{48}, and \iso{Bi}{209} are larger.
The larger valence spaces require larger unitary transformations to decouple,
leading to larger $\Omega_G$ in the hunter-gatherer approach.
This changes the convergence of the BCH expansion~\eqref{eq:BCHExpansion},
leading to larger differences between the schemes.

On the other hand, I find no obvious dependence model space size.
Calculations of \iso{Ca}{48} performed in model spaces of 7 major harmonic oscillator shells
show similar differences between predictions obtained using the flow equation, split Magnus, and hunter-gatherer approaches as the calculations in 13 major oscillator shells in Fig.~\ref{fig:Ca48Comparison}.

In Table~\ref{tab:Bi209Table},
I compare the energies of selected states in \iso{Bi}{209} with predictions from coupled-cluster theory~\cite{Hagen2014RPP_CCReview, Bonaiti2025arxiv_Pb266}.
The employed approximation, PA-EOM-CCSD(3p-2h)$_\mathrm{pert}$, has been shown to perform similarly to the VS-IMSRG(2)~\cite{Morris2018PRL_Sn100}.
The coupled-cluster results are generally closer to the split Magnus results than the hunter-gatherer results.
However, the differences between the hunter-gatherer and split Magnus approaches are comparable in size to the differences between coupled-cluster and the split Magnus approaches, not larger.
Thus the observed differences may be seen as being reflective of the overall uncertainty of the IMSRG(2) truncation.

\prlsection{Conclusion}
I studied the hunter-gatherer scheme for solving the IMSRG(2) equations,
which underlies the improved IMSRG(3f$_2$) approximation,
seeking to understand differences between this scheme and standard approaches (direct integration of the flow equation and a split Magnus expansion scheme).
I find that the hunter-gatherer scheme is very reliable for single-reference IMSRG(2) calculations,
differing very little from standard schemes.
For valence-space IMSRG(2) calculations,
differences between the hunter-gatherer and split Magnus schemes are larger.
In some cases, these differences are comparable to other uncertainties related to the IMSRG(2) approximation,
e.g., differences between IMSRG(2) and VS-IMSRG(2) calculations for \iso{Ca}{48}.
These differences are similar in size to expected IMSRG(3) corrections,
especially for calculations involving larger valence spaces.
The IMSRG(3f$_2$) remains valuable for more precise calculations and uncertainty quantification,
but the uncertainty of the hunter-gatherer scheme should be quantified when it is employed.
Improved approximations are expected reduce this uncertainty,
as it is related to the normal-ordered two-body truncation of the IMSRG(2) commutators.

\prlsection{Acknowledgments}
I thank Bingcheng He and Ragnar Stroberg for insightful discussions on the hunter-gatherer scheme and the IMSRG(3f$_2$) approximation
and Francesca Bonaiti for discussions on coupled-cluster computations of \iso{Bi}{209}.
All calculations were performed using the \texttt{imsrg++} code, build version \texttt{\detokenize{devel_fda158a}}~\cite{Stroberg2026_IMSRGGit}.
This work was supported 
by the Laboratory Directed Research and Development Program of Oak Ridge National Laboratory, managed by UT-Battelle, LLC, for the U.S.\ Department of Energy
and by the U.S.\ Department of Energy, Office of Science, Office of Advanced Scientific Computing Research and Office of Nuclear Physics, Scientific Discovery through Advanced Computing (SciDAC) program (SciDAC-5 NUCLEI).
This research used resources of the Oak Ridge Leadership Computing Facility located at Oak Ridge National Laboratory, which is supported by the Office of Science of the Department of Energy under contract No.~DE-AC05-00OR22725.
The authors gratefully acknowledge the Gauss Centre for Supercomputing e.V.\ (www.gauss-centre.eu) for funding this project by providing computing time through the John von Neumann Institute for Computing (NIC) on the GCS Supercomputer JUWELS at Jülich Supercomputing Centre (JSC).

This manuscript has been authored by UT-Battelle, LLC, under contract DE-AC05-00OR22725 with the US Department of Energy (DOE). The US government retains and the publisher, by accepting the article for publication, acknowledges that the US government retains a nonexclusive, paid-up, irrevocable, worldwide license to publish or reproduce the published form of this manuscript, or allow others to do so, for US government purposes. DOE will provide public access to these results of federally sponsored research in accordance with the DOE Public Access Plan (\url{https://www.energy.gov/doe-public-access-plan}).


\prlsection{Data availability}
The data supporting this work is publicly available~\cite{heinz_2026_18341411}.

\bibliography{ref}

\end{document}